%% file: Trellis.tex
\documentclass[prl,twocolumn,superscriptaddress]{revtex4}

\usepackage{hyperref}
\usepackage{amsfonts,amsmath, amsthm, amssymb}
\usepackage{graphicx}
\usepackage{color}
\usepackage[noend]{algorithmic}

\newcommand{\ket}[1]{\lvert #1 \rangle} 
\newcommand{\bra}[1]{\langle #1 \rvert}

\newcommand{\gf}[1]{\mathbb{F}_{#1}}       
\newcommand{\gr}[1]{\mathsf{#1}}  
\newcommand{\prob}{\mathrm{Pr}}
\newcommand{\Sperp}{S^\perp}
\newcommand{\Ftr}[1]{C_{#1}^f}
\newcommand{\Pst}[1]{C_{#1}^p}
\newcommand{\St}[1]{S_{\mbox{\scriptsize start} \leq #1}}
\newcommand{\End}[1]{S_{\mbox{\scriptsize end} \leq #1}}

\renewcommand{\section}[1]{\textbf{#1.}}

\DeclareMathOperator{\eqdef}{\triangleq}

\newtheorem{lemma}{Lemma}
\newtheorem{theorem}{Theorem}

\begin{document}

\title{Trellises for stabilizer codes: definition and uses}

\author{Harold Ollivier} 
\affiliation{Perimeter Institute, 31 Caroline St. N, Waterloo, ON N2L
2Y5, Canada.}  
\author{Jean-Pierre Tillich}
\affiliation{INRIA, Projet Codes, Domaine de Voluceau BP 105, F-78153
Le Chesnay cedex, France.}

\date{}

\begin{abstract}
Trellises play an important theoretical and practical role for
classical codes. Their main utility is to devise complexity-efficient
error estimation algorithms. Here, we describe trellis representations
for quantum stabilizer codes. We show that they share the same
properties as their classical analogs. In particular, for any
stabilizer code it is possible to find a minimal trellis
representation. Our construction is illustrated by two fundamental
error estimation algorithms.
\end{abstract}

\maketitle

\section{Introduction}
Since the discovery of efficient quantum algorithms for solving hard
classical problems, many efforts have been devoted to building quantum
processing devices. While small scale prototypes are readily
available, scalability remains a practical issue because of the
extreme sensibility of quantum devices to external noise. Fortunately,
theoretical advances, such as the discovery of error correction
schemes and fault-tolerant implementations, have notably cleared the
future of quantum computing. One way of building quantum codes is
through the stabilizer formalism \cite{Got97a, CRSS98a}. An $(n,k)$
stabilizer code protects $k$ qubits by encoding them into an $n$-qubit
register. The error recovery procedure involves the measurement of a
syndrome (i.e., a vector in $\gf{2}^{n-k}$) which is used to partially
discriminate the actual error $E$ from all possible ones. The role of
the error model, and in general of any a priori information about $E$,
is to allow further discrimination in order to find the most likely
guess $\hat E$. While one can imagine computing the likelihood of all
possible errors compatible with the measured syndrome, such method is
impractical when codes are large. This is because there are of order
$2^{n+k}$ such elements. This problem is well known in classical
coding theory, but was rightfully ignored in the quantum case as block
codes that can be physically implemented have extremely small
length. However, with the development of quantum communication and the
advent of other coding strategies, this shall no longer be the case.

In classical coding theory, one often relies on a graphical
representation of the code, called a {\em trellis}, to perform error
estimation. For instance, trellises yield many complexity-efficient
error estimation schemes for memoryless channels as well as means of
estimating the noise parameters. In particular, it can be used to
calculate with linear complexity the most likely error for a
convolutional code of bounded memory over any memoryless channel.  In
this article, we apply general results concerning group codes for
classical communication~\cite{FT93a} to show that a similar
representation is available for quantum stabilizer codes. Two error
estimation schemes that exploit trellises are introduced for
memoryless channels. We show that the complexity of these algorithms
is related to the number of trellis vertices, and provide a
construction of a trellis which minimizes this quantity. One of these
algorithms achieves the performance of an algorithm for convolutional
codes that was proposed in \cite{OT03a}, but with a significantly
lower complexity. 

\section{Stabilizer codes: elementary facts}\label{sec:stabilizer}
In the rest of this paper, some familiarity with quantum computation
is assumed. This section provides a brief introduction to stabilizer
codes, for a more detailed introduction the reader is redirected
to~\cite{Got97a, CRSS98a, NC00a} and references therein.

In what follows, without loss of generality, the quantum register of
interest has $n$ physical qubits.
\paragraph{Preliminaries.}
Stabilizer codes rely heavily on properties of $\mathcal G_n$, the
$n$-qubit Pauli group. This group is defined in terms of the Pauli
matrices for a single qubit: $\mathcal I = \bigl
(\begin{smallmatrix}1&0 \\ 0&1\end{smallmatrix}\bigr ), \mathcal X =
\bigl(\begin{smallmatrix}0&1 \\ 1&0 \end{smallmatrix}\bigr ), \mathcal
Y = \bigl (\begin{smallmatrix}0&- i \\ i&0 \end{smallmatrix}\bigr ),
\mathcal Z = \bigl(\begin{smallmatrix}1&0 \\
0&-1\end{smallmatrix}\bigr)$. The group ${\mathcal G}_n$ is the
multiplicative group generated by the $n$-fold tensor products of
single qubit Pauli matrices.

For our purpose here, phases are irrelevant and it will be more
convenient to work with the effective Pauli group $G_n \eqdef \mathcal
G_n / \{\pm \mathcal I^{\otimes n} ; \pm i \mathcal I^{\otimes n}\}$
(see~\cite{CRSS98a}). The elements of $G_1$ will be denoted by
$I\eqdef [\mathcal I]$, $X\eqdef [\mathcal X]$, $Y\eqdef [\mathcal
Y]$, and $Z\eqdef [\mathcal Z]$.  Here, $[\mathcal P]$ denotes the
equivalence class of $\mathcal P \in \mathcal G_n$, that is $\{\pm
\mathcal P,\pm i \mathcal P\}$. Note that $G_n$ is Abelian, so that we
will use the additive notation for its group operation. Since $G_n
\cong G_1^n$, we often view $P \in G_n$ as an $n$-tuple $(P^i)_{i =
1}^n$ with entries in $G_1$.

The crucial fact about $\mathcal G_n$, is that any pair of elements
$\mathcal P,\mathcal Q$ either commutes or anti-commutes. This leads
to the definition of an inner product ``$\star$'' for elements of
$G_n$ such that $(P^i)_i \star (Q^i)_i = \sum_i P^i \star Q^i \mod
2$. Here, $P^i \star Q^i = 1$ if $P^i \neq Q^i$, $P^i \neq I$ and $Q^i
\neq I$; and $P^i \star Q^i = 0$ otherwise. One can then check easily
that $\mathcal P, \mathcal Q \in \mathcal G_n $ commute if and only if
$[\mathcal P]\star [\mathcal Q] = 0$.

\paragraph{Error model.}
Stabilizer codes can accommodate for a broad class of channels. For
simplicity, only memoryless Pauli channels will be considered,
although the tools presented in this paper extend to other memoryless
channels. Memoryless Pauli channels act on the whole $n$-qubit
register as $\sigma \rightarrow \Psi(\sigma) = (\Psi^1 \otimes \Psi^2
\otimes \ldots \otimes \Psi^n) (\sigma)$. Above $\Psi^i$ is a
$1$-qubit channel whose action on $\rho$, a single qubit density
operator, is given by $ \Psi^i(\rho) = \sum_{\mathcal E\in\{\mathcal
I,\mathcal X,\mathcal Y,\mathcal Z\}} \prob_i([\mathcal E]) \mathcal
E\rho \mathcal E$, with $\prob_i(\cdot)$ a probability distribution
over $G_1$.

\paragraph{Definition of the code subspace.}
The code subspace $C$ of an $(n,k)$ stabilizer code is the largest
subspace stabilized by the action of $\mathcal S$, an Abelian subgroup
of $\mathcal G_n$. For the code to protect $k$ qubits using $n$,
i.e. to be of rate $k/n$, $\mathcal S$ must be generated by $n-k$
independent operators $\mathcal S_j$, and be such that $-\mathcal
I^{\otimes n} \notin \mathcal S$. The code subspace is equivalently
defined by $n-k$ eigenvalue equations: $\ket \psi \in C$ if and
only if $\forall j \in \{1,\ldots,n-k\}, \ \mathcal S_j \ket \psi =
\ket \psi$.

To study the main properties of these codes, phases are again
irrelevant. More precisely, it is sufficient to represent the set of
generators of the stabilizer group $\{\mathcal S_j\}_j$ by the set of
equivalence classes $\{S_j\}$ where $S_j = [\mathcal S_j]$ which
generate a subgroup $S$ of $G_n$. Using a slight abuse in terminology,
we also call $S$ the stabilizer group of the code and $\{S_j\}_j$ the
stabilizer set of the code.

\paragraph{Quantum convolutional codes.}
Following the definition of~\cite{OT03a,OT04a}, an $(n,k)$ stabilizer
code is convolutional with parameters $(\eta,\kappa)$ if there exists
a set of generators $\{S_j\}_j$ of its stabilizer group $S$ with an
$\eta$-qubit shift invariance property. More precisely, the values
$S_j^i$ must be equal to the entries $H_j^i \in G_1$ of an infinite
matrix $H$ which satisfies $H_{j+\eta - \kappa}^{i+\eta} =
H_j^i$ for every $i,j$.

For a convolutional code given by the stabilizer set $\{S_j\}_j$, the
{\em memory} is defined as $m = \max_i \sum_j \sharp(S_j^i)$. Here,
$\sharp(S_j^i) = 0$ if either $S_j^k = I$ for all $k > i$ or $S_j^k =
I$ for all $k\leq i$; and $\sharp(S_j^i) = 1$ otherwise. It is
important to note that the memory depends on the set $\{S_j\}_j$ and
is thus not an intrinsic property of the code.

\paragraph{Error estimation.}
The goal of error estimation is to infer channel errors from their
action on the state of the quantum register. In the context of
stabilizer codes, the necessary information is provided by the
measurement of the Hermitian operators $\mathcal S_j$.

Let $\mathcal E\in \mathcal G_n$ be the actual, yet unknown, quantum
error that affected the state $\ket \psi \in C$. The measurement of
the operators $\mathcal S_j$ on $\ket{\psi'}\eqdef \mathcal E\ket
\psi$ defines a binary vector of length $n-k$ called the syndrome of
$\mathcal E$: $s(\mathcal E) \eqdef (s^j(\mathcal E))_{j=1}^{n-k}
\eqdef \frac{1}{2}(1 - \bra{\psi'} \mathcal S_j \ket{\psi'}) = (S_j
\star [\mathcal E])_{j = 1}^{n-k}$. Among all possible error operators
$\mathcal F\in \mathcal G_n$, only some of them are compatible with
$s(\mathcal E)$ (i.e., they satisfy $s(\mathcal F) = s(\mathcal
E)$). Therefore, they belong to a coset of $N(\mathcal S)$, the
normalizer of $\mathcal S$ in $\mathcal G_n$. Equivalently, these
compatible errors $\mathcal F$ are such that $[\mathcal F]$ belongs to
a coset of $\Sperp \eqdef \{ P \in G_n : P\star Q = 0, \ \forall Q \in
S\}$. While, the knowledge of the syndrome restricts the class of
errors that could have happened during the transmission, the error
model further discriminates between these elements by assigning them
probabilities. Error recovery then uses these probabilities to find a
best guess $\hat{\mathcal E}$ for the actual error $\mathcal E$. For
instance, maximum likelihood error estimation consists in finding a
most likely $\hat{\mathcal E}$ compatible with the measured syndrome
$s(\mathcal E)$. Trellises are both aimed at computing these
probabilities and at choosing a best guess efficiently.
 
\section{Trellises for stabilizer codes}\label{sec:trellis}
Considering the previous remarks about error estimation, it is natural
to seek a representation of cosets of $\Sperp$ in which it is easy to
search for their most likely element. This is the main motivation for
the definition of trellises for quantum as well as for classical
codes. In a broader context, this motivation extends to the
calculation of quantities on which error estimation is based, e.g.\@
the likelihood function, the a posteriori qubit error probability,
etc.

\paragraph{Definition.}
In the rest of this section, $S$ denotes the stabilizer group of a
quantum code with parameters $(n,k)$, and $\{S_j\}_{j=1}^{n-k}$ its
stabilizer set.

An {\em $n$-section trellis} relative to the stabilizer set
$\{S_j\}_{j=1}^{n-k}$ and syndrome $s \in \gf 2^{n-k}$ is a directed
graph with the following properties:

(1) Its vertices can be grouped into $n+1$ sets $\gr{V}_i$, with
$|\gr{V}_0| = |\gr{V}_n|= 1$. The set $\gr V_i$ is called the $i$-th
{\em state space} of the trellis

(2) Its edges are directed and can be grouped into $n$ sets
$\gr{E}_i$.  An edge $\gr{e} \eqdef (\gr{v}\gr{w})$ is said to be
issued at vertex $\gr{v}$ and ending at vertex $\gr{w}$. Edges in
$\gr{E}_i$ are issued from a vertex of $\gr{V}_{i-1}$ and end at a
vertex of $\gr{V}_i$. The set $\gr E_i$ is called the $i$-th {\em section}
of the trellis.

(3) An edge $\gr{e}\in \gr{E}_i$ bears a label $l(\gr{e})$ such that
$l(\gr{e}) \in G_1$. We say that $l(\gr e)$ is the {\em Pauli-label}
of $\gr e$. 

(4) Each element $P \in {G}_n$ with syndrome $s$ is associated to a
unique directed path $(\gr{e}^1, \ldots, \gr{e}^n)$ such that
$l(\gr{e}^i) = P^i$.

\begin{figure}[tbp] 
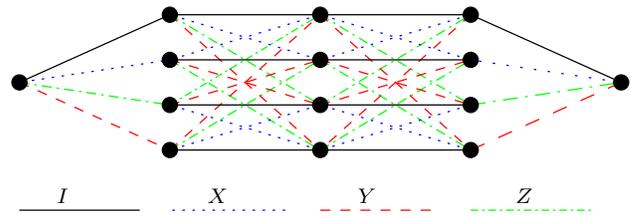
\caption{Trellis representation for the $4$-qubit code and syndrome
$(0,0)$. Here, $|\gr V_0|=|\gr V_4|=1$ and $|\gr V_1|=|\gr V_2|=|\gr
V_3|=4$.}\label{fig:example1}
\end{figure}

An example of trellis is considered in Figure~\ref{fig:example1} for
the $4$-qubit code with stabilizer group generated by $\{XXXX,ZZZZ\}$.

\paragraph{Construction.}
Given a code stabilized by $S$, there are many possible trellises
representing the coset of $\Sperp$ of syndrome $s$.  We provide here a
simple construction for which the number of vertices in each $\gr V_i$
is bounded by $2^{n-k}$. In analogy with classical codes, this trellis
will be called the {\em Wolff trellis} of the code defined by the
stabilizer set $\{S_j\}_j$ relative to the syndrome $s$.

For every $i \in \{0,\dots,n\}$, let $\pi_i$ be a mapping from $G_n$
to itself defined by $\pi_i(P^1,P^2,\dots,P^n) =
(P^1,\dots,P^i,I,\dots,I)$ (with the convention that
$\pi_0(P^1,P^2,\dots,P^n) = (I,\dots,I)$).  Let $P_s$ be an arbitrary,
but fixed, element of $G_n$ with syndrome $s$.  For every $i \in
\{0,\dots,n\}$, $\gr V_i$ is a subset of $\gf2^{n-k}$ defined by $\{
(S_j \star \pi_i(P))_{j=1}^{n-k}:P \in P_s + S^\perp\}$. A vertex $\gr
v \in \gr V_i$ is connected to vertex $\gr w \in \gr V_{i+1}$ with an
edge labelled by $E$ iff there exists $P \in P_s + S^\perp$ such that
$\gr v = (S_j \star \pi_i(P))_{j=1}^{n-k}$, $\gr w = (S_j \star
\pi_{i+1}(P)_{j=1}^{n-k}$ and $P^{i+1}=E$.

\begin{figure*}[tbp] 
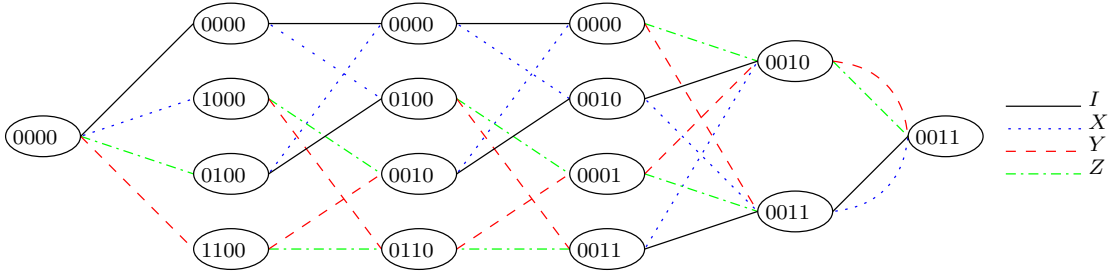
\caption{Trellis for the $5$-qubit code defined by the stabilizer set
$\{ZXIII,XZXII,IXZXI,IIXZX\}$ and for the syndrome $s = (0011)$
obtained through the second construction. Here $|\gr V_0|=|\gr V_5|=1$
and $|\gr V_1|=|\gr V_2|=|\gr V_3|=4$ and $|\gr
V_4|=2$.}\label{fig:example2}
\end{figure*}

An exemple of trellis obtained in this way is given in
Figure~\ref{fig:example2} for the $5$-qubit code associated to the
stabilizer set $\{ZXIII,XZXII,IXZXI,IIXZX\}$.

As hinted by this example, the Wolff trellis of convolutional codes
has state-spaces that are bounded in size by $2^m$, where $m$ is the
memory associated to the chosen stabilizer set $\{S_j\}_j$. Note that
this bound is achieved for all indices $i$ such that $m = \sum_j
\sharp(S_j^i)$.

\paragraph{Minimality.}
As it will be seen below, the complexity of many useful algorithms
using trellises is linear in their number of vertices. This raises the
issue of finding a trellis which minimizes this quantity. If one is
willing to change the stabilizer set for the code in order to put it
in a {\em trellis oriented form}, then the Wolff trellis is
minimal. Indeed, the trellis obtained in this way does not only
minimize the number of vertices but also the {\em state space
profile}. As for classical trellises, we define the state profile of
the trellis of an $(n,k)$ code by the $(n+1)$-tuple
$(\xi_0,\xi_1,\dots,\xi_n)$ where $\xi_i = \log_2 |\gr V_i|$ where
$\gr V_i$ is the $i$-th state space of the trellis. In other words, we
are going to prove that the Wolff trellis applied to a stabilizer set
in trellis oriented form minimizes each $\xi_i$ individually. Without
loss of generality, we now assume that the trellis is associated to
the syndrome $s = (0, \ldots 0)$.

First, we define the trellis oriented form of a stabilizer set. For
$1\leq j\leq n-k$, let $c(j)$ and $d(j)$ be respectively the position
of the first (resp. last) component of $S_j$ which is different from
$I$.  We say that the stabilizer set $\{S_j\}$ is in trellis oriented
form if and only if for all $j$ (1) $S_{j'}^{c(j)} = I$ for $j' > j+1$
and $S_{j+1}^{c(j)} \neq S_j^{c(j)}$; and (2) there is at most one
$j'\neq j$ such that $b(j) = b(j')$ and in such case $S_j^{b(j)} \neq
S_{j'}^{b(j')}$. Note that any stabilizer set $\{S_j\}$ can be put in
trellis oriented form by indices permutations and group
additions. 

Second, we show a lower bound on $\xi_i$. For an $(n,k)$ çstabilizer
code given by the stabilizer set $\{S_j\}_j$, and for $i \in
\{0,\dots, n\}$, let $\Ftr i$ (the {\em future} subgroup) be the
subgroup of $\Sperp$ whose elements have their first $i$ components
equal to $I$; and let $\Pst i$ (the {\em past} subgroup) be the
subgroup of $\Sperp$ whose elements have their last $n-i$ components
equal to $I$.  We then have the following lemma.
\begin{lemma}\rm 
$\xi_i \geq n+k - \log_2 |C_i^p| - \log_2 |C_i^f|$.
\end{lemma}

\begin{proof}
Let $\gr v \in \gr V_i$, and $C_{\gr v}$ be the set of elements of
$\Sperp$ that correspond to a path in the trellis associated to the
all-zero syndrome and passing through $\gr v$. Let $\gr P$ and $\gr F$
be the set of all paths that go from $\gr v_0$ to $\gr v$ (resp. from
$\gr v$ to $\gr v_n$). Let $Q$ be a fixed element of $C_{\gr v}$ and
$\gr q_{\gr P} \gr q_{\gr F}$ its corresponding path in the trellis,
where $\gr q_{\gr P} \in \gr P$ and $\gr q_{\gr F} \in \gr F$. By
construction of the trellis, we have $|C_{\gr v}| = |\gr P||\gr
F|$. Note that any path $\gr p \in \gr P$ can be extended into a path
$\gr p \gr q_{\gr F}$ from $\gr v_0$ to $\gr v_n$. Since any path $\gr
p \gr q_{\gr F}$ can be associated to an element of $Q + \Pst i$, we
have $|\gr P| \leq |Q + \Pst i| = |\Pst i|$. Similarly $|\gr F| \leq
|\Ftr i|$, which yields $2^{n+k} = |\Sperp| = \sum_{\gr v \in \gr V_i}
|C_{\gr v}| \leq |\gr V_i||\Pst i||\Ftr i|$. This gives the desired
bound on $\xi_i$ by taking the logarithm.
\end{proof}

Finally we conclude with the following theorem.
\begin{theorem}
The Wolff trellis achieves the previous bound on $\xi_i$ for each
$i$ when the stabilizer set is in trellis oriented form.
\end{theorem}

\begin{proof}  
Let $\St{i}$ be the subset of operators of $\{S_j\}$ which have at
least one of their $i$-first components different from $I$. Let
$\End{i}$ be the subset of operators of $\{S_j\}$ which have all their
last $n-i$ components equal to $I$. It is straightforward to check
that $\xi_i = |\St i| - |\End i|$. Note that, $\Pst i$ is the
subgroup of $G_n$ with $I$ on their $n-i$ last components and
orthogonal to all elements of $\St i$. This implies $\log_2 |\Pst i| =
2i - |\St i|$. Using a similar argument, we get that $\log_2 |\Ftr i|
= 2(n-i) - (n-k - |\End i|)$. Adding these equalities, we conclude
that $\xi_i + \log_2 |\Pst i| + \log_2 |\Ftr i| = n+k$.
\end{proof}

\section{Using trellises of stabilizer codes}

\paragraph{Min-Sum (Viterbi) algorithm.}
The Min-Sum algorithm is certainly one of the most widely employed
algorithm that benefits from the trellis representation of classical
codes. Here, we present a Min-Sum algorithm for stabilizer codes that
computes the most likely error for memoryless Pauli channels given a
measured syndrome $s$ by using a trellis associated to the code.

Consider an $(n,k)$ stabilizer code with stabilizer set $\{S_j\}_j$.
Define the likelihood of $P \in G_n$ as $\sum_{i=1}^n \log
\prob_i(P^i)$. Consider the $n$-section trellis for this quantum code
associated to the syndrome $s$. The naming conventions for the
vertices and edges are set as in previous sections. For each edge
$\gr{e}^i$ of $\gr{E}_i$, define its weight $wt(\gr{e}^i) = - \log
\prob_i(l(\gr{e}^i))$. By construction, the sum of weights along the
path in the trellis that represents $P$ is equal to the opposite of
the likelihood. The task which consists in finding a most likely error
$\hat E\in G_n$ with syndrome $s$ is thus equivalent to finding a
lowest weight path $(\hat{\gr e}^1, \ldots, \hat{\gr e}^n)$ in the
trellis associated to $s$.

This can be done by constructing recursively some sets $\gr C_i$ of
lowest weight error candidates. More precisely, $\gr C_i$ contains
couples $(\gr{c},w)$ where $\gr{c}$ is a path issued from $\gr{v}_0$
that ends on a vertex of $\gr{V}_i$ and where $w = wt(\gr c)$.

\noindent\begin{tabular}{|@{}p{0.2cm}@{}p{0.2cm}@{}p{\columnwidth - 0.46cm}|}
\hline\hline
\multicolumn{3}{|c|}{Min-Sum Algorithm} \\ 
\hline
\multicolumn{3}{|l|}{{\em Initialization:} $\gr C_0 := \{(\gr v_0,0)\}$ 
and $\gr C_i := \emptyset$ for $i\leq 1$} \\ 
\multicolumn{3}{|l|}{\em Main step:} \\
\multicolumn{3}{|l|}{\textbf{for} $i$ from 1 to $n$ \textbf{do}} \\ 
& \multicolumn{2}{l|}{\textbf{for all} $\gr v \in \gr V_i$ \textbf{do}} \\
&& {Put in $\gr C_i$ the pair 
$(\gr c',wt(\gr c'))$, where $\gr c'$ is a path of minimum weight 
(ties are broken at random) among all paths that: (1) end in $\gr v$; 
(2) have their $i-1$ first vertices given by a path $\gr c$ of 
$\gr C_{i-1}$.} \\ \hline
\end{tabular}
Note that the time complexity of this algorithm is linear in the
number of vertices in the trellis.

\paragraph{Sum-Product algorithm.}
While the Min-Sum finds a most likely error compatible with the
observed syndrome $s$, the Sum-Product aims at calculating marginal
error probabilities for physical qubits. That is, $p_i(P) \eqdef
\prob(\mbox{error at qubit } i = P | s)$ where $P \in G_1$. By
definition of the trellis, this probability is equal to the
probability that a path $(\gr e^i)_{i=1}^n$ from $\gr v_0$ to $\gr
v_n$ is such that $l(\gr e^i) = P$.

The Sum-Product algorithm computes for each vertex $\gr v$ of the
trellis associated with $s$ a ``forward'' probability $f(\gr v)$ and a
``backward'' probability $b(\gr v)$. Both are then used to calculate
the marginal probabilities $p_i(P)$.

\noindent\begin{tabular}{|@{}p{0.2cm}@{}p{0.2cm}@{}p{\columnwidth - 0.55cm}|}
\hline\hline
\multicolumn{3}{|c|}{Sum-Product Algorithm} \\ 
\hline
\multicolumn{3}{|p{\columnwidth - 0.15cm}|}{{\em Initialization:} $f(\gr v_0) := 1$ and $b( \gr v_n):=1$; and $f(\gr v) := 0$ and $b(\gr v) := 0$ for all other vertices}\\ 
\multicolumn{3}{|l|}{\em Forward pass :} \\
\multicolumn{3}{|l|}{\textbf{for} $i=1$ to $n$ \textbf{do}} \\
&\multicolumn{2}{l|}{\textbf{for all} $\gr v \in \gr V_i$ \textbf{do} $f(\gr v ) := \sum_{\gr w \in \gr V_i^{-}(\gr v)} f(\gr w) \prob_i(l(\gr w \gr v))$} \\
&\multicolumn{2}{l|}{$ F_i := \sum_{\gr v \in \gr V_{i}} f(\gr v)$} \\
&\multicolumn{2}{l|}{\textbf{for all} $\gr v \in \gr V_i$ \textbf{do} $f(\gr v) := f(\gr v)/F_i $} \\
\multicolumn{3}{|l|}{\em Backward pass :} \\
\multicolumn{3}{|l|}{\textbf{for} $i=n-1$ down to $0$ \textbf{do}} \\
&\multicolumn{2}{l|}{\textbf{for all} $\gr v \in \gr V_i$ \textbf{do} $b(\gr v) = \sum_{\gr w \in \gr V_i^{+}(\gr v)} b(\gr w) \prob_{i+1}(l(\gr v \gr w))$} \\
&\multicolumn{2}{l|}{$ B_i := \sum_{\gr v \in \gr V_{i}} b(\gr v)$} \\
&\multicolumn{2}{l|}{\textbf{for all} $\gr v \in \gr V_i$ \textbf{do} $b(\gr v) := b(\gr v)/B_i $}  \\ 
\multicolumn{3}{|l|}{\em Final pass :} \\
\multicolumn{3}{|l|}{\textbf{for} $i=1$ to $n$ \textbf{do}} \\
&\multicolumn{2}{l|}{\textbf{for all} $P \in G_1$ \textbf{do}} \\ 
&&{$p_i(P) := \sum_{\gr v \gr w \in \gr E_i(P)} f(\gr v) b(\gr w) \prob_i(P)$} \\ \hline
\end{tabular}
Above, (1) $\gr V_i^{-}(\gr v)$ is the set of vertices $\gr w$ in $\gr
V_{i-1}$ which are adjacent to $\gr v$; (2) $\gr V_i^{+}(\gr v)$ is
the set of vertices $\gr w$ in $\gr V_{i+1}$ which are adjacent to
$\gr v$; and (3) $\gr E_i(P)$ is the set of edges between $\gr
V_{i-1}$ and $\gr V_i$ that bear the Pauli-label $P$.

Once again, the practical relevance of this algorithm is due to the
fact that its complexity is linear in the number of vertices in the
trellis.

\paragraph{Computing the weight enumerator polynomial.}
The {\em weight enumerator polynomial} is a trivariate polynomial
given by $A(x,y,z) \eqdef \sum_{1 \leq u,v,w \leq n} a_{u,v,w}x^u y^v
z^w$, where $a_{u,v,w} \eqdef |\{P \in N(S):
|P|_X=u,|P|_Y=v,|P|_Z=w\}|$ and where $|P|_E$ denotes the number of
coordinates of $P$ which are equal to $E$. It is possible to extract
from $A(x,y,z)$ a lot of useful information, e.g.\@ bounds on the
fidelity of the recovered state after decoding~\cite{DSS98a}. As for
classical codes, the weight enumerator can be computed with complexity
linear in the number of vertices of the trellis. For this purpose
intermediate polynomials $A_{\gr v} (x,y,z)$ are calculated for each
vertex of the trellis associated to $s = (0,\ldots 0)$. Also define
the polynomials $Q_I(x,y,z) \eqdef 1$, $Q_X(x,y,z) \eqdef x$,
$Q_Y(x,y,z) \eqdef y$ and $Q_Z(x,y,z) \eqdef z$.

\noindent\begin{tabular}{|@{}p{0.2cm}@{}p{0.2cm}@{}p{\columnwidth - 0.46cm}|}
\hline\hline
\multicolumn{3}{|c|}{Computation of $A(x,y,z)$} \\ 
\hline
\multicolumn{3}{|l|}{{\em Initialization:} $A_{\gr v_0}(x,y,z) = 1$} \\ 
\multicolumn{3}{|l|}{\em Main step:} \\
\multicolumn{3}{|l|}{\textbf{for} $i$ from 1 to $n$ \textbf{do}} \\ 
& \multicolumn{2}{l|}{\textbf{for all} $\gr v \in \gr V_i$ \textbf{do}} \\
&& $A_{\gr v}(x,y,z) := \sum_{\gr w \in \gr V^{+}_i(\gr v)} A_{\gr
w}(x,y,z) Q_{l(\gr w\gr v)}(x,y,z)$ \\ 
\multicolumn{3}{|l|}{$A(x,y,z) := A_{\gr v_n}(x,y,z)$} \\
\hline 
\end{tabular}
Above, $\gr V_i^{-}(\gr v)$ is the set of vertices $\gr w$ in $\gr
V_{i-1}$ which are adjacent to $\gr v$.

The proof of correctness for these three algorithms follows from the
correctness of the more general Min-Sum and Sum-Product algorithms
presented in \cite{KFL01a}.

\section{Conclusion}
From a practical point of view, we have proposed a definition for the
trellis of a stabilizer code together with two constructions and three
algorithms that take advantage of this representation. Following the
same path, several algorithms for classical codes running on trellises
can be generalized to the quantum case. Among them, estimation of
noise parameters (or equalization) seems a promising avenue for
enhancing the performance of quantum optics fiber communications using
near-future quantum technology.


\end{document}

%% file: exemple-a.pstex_t
\begin{picture}(0,0)%
\includegraphics{exemple-a}%
\end{picture}%
\setlength{\unitlength}{4144sp}%
\begingroup\makeatletter\ifx\SetFigFont\undefined%
\gdef\SetFigFont#1#2#3#4#5{%
  \reset@font\fontsize{#1}{#2pt}%
  \fontfamily{#3}\fontseries{#4}\fontshape{#5}%
  \selectfont}%
\fi\endgroup%
\begin{picture}(3706,1235)(3638,-4393)
\put(4816,-4336){\makebox(0,0)[lb]{\smash{{\SetFigFont{8}{9.6}{\familydefault}{\mddefault}{\updefault}{\color[rgb]{0,0,0}$X$}%
}}}}
\put(5716,-4336){\makebox(0,0)[lb]{\smash{{\SetFigFont{8}{9.6}{\familydefault}{\mddefault}{\updefault}{\color[rgb]{0,0,0}$Y$}%
}}}}
\put(6661,-4336){\makebox(0,0)[lb]{\smash{{\SetFigFont{8}{9.6}{\familydefault}{\mddefault}{\updefault}{\color[rgb]{0,0,0}$Z$}%
}}}}
\put(3916,-4336){\makebox(0,0)[lb]{\smash{{\SetFigFont{8}{9.6}{\familydefault}{\mddefault}{\updefault}{\color[rgb]{0,0,0}$I$}%
}}}}
\end{picture}%

%% file: exemple-c.pstex_t
\begin{picture}(0,0)%
\includegraphics{exemple-c}%
\end{picture}%
\setlength{\unitlength}{4144sp}%
\begingroup\makeatletter\ifx\SetFigFont\undefined%
\gdef\SetFigFont#1#2#3#4#5{%
  \reset@font\fontsize{#1}{#2pt}%
  \fontfamily{#3}\fontseries{#4}\fontshape{#5}%
  \selectfont}%
\fi\endgroup%
\begin{picture}(6702,1608)(-231,-1315)
\put(5221,-556){\makebox(0,0)[lb]{\smash{{\SetFigFont{8}{9.6}{\familydefault}{\mddefault}{\updefault}{\color[rgb]{0,0,0}$0011$}%
}}}}
\put(946,-781){\makebox(0,0)[lb]{\smash{{\SetFigFont{8}{9.6}{\familydefault}{\mddefault}{\updefault}{\color[rgb]{0,0,0}$0100$}%
}}}}
\put(946,-331){\makebox(0,0)[lb]{\smash{{\SetFigFont{8}{9.6}{\familydefault}{\mddefault}{\updefault}{\color[rgb]{0,0,0}$1000$}%
}}}}
\put(946,119){\makebox(0,0)[lb]{\smash{{\SetFigFont{8}{9.6}{\familydefault}{\mddefault}{\updefault}{\color[rgb]{0,0,0}$0000$}%
}}}}
\put(946,-1231){\makebox(0,0)[lb]{\smash{{\SetFigFont{8}{9.6}{\familydefault}{\mddefault}{\updefault}{\color[rgb]{0,0,0}$1100$}%
}}}}
\put(2071,-1231){\makebox(0,0)[lb]{\smash{{\SetFigFont{8}{9.6}{\familydefault}{\mddefault}{\updefault}{\color[rgb]{0,0,0}$0110$}%
}}}}
\put(2071,-781){\makebox(0,0)[lb]{\smash{{\SetFigFont{8}{9.6}{\familydefault}{\mddefault}{\updefault}{\color[rgb]{0,0,0}$0010$}%
}}}}
\put(2071,-331){\makebox(0,0)[lb]{\smash{{\SetFigFont{8}{9.6}{\familydefault}{\mddefault}{\updefault}{\color[rgb]{0,0,0}$0100$}%
}}}}
\put(2071,119){\makebox(0,0)[lb]{\smash{{\SetFigFont{8}{9.6}{\familydefault}{\mddefault}{\updefault}{\color[rgb]{0,0,0}$0000$}%
}}}}
\put(3196,119){\makebox(0,0)[lb]{\smash{{\SetFigFont{8}{9.6}{\familydefault}{\mddefault}{\updefault}{\color[rgb]{0,0,0}$0000$}%
}}}}
\put(3196,-331){\makebox(0,0)[lb]{\smash{{\SetFigFont{8}{9.6}{\familydefault}{\mddefault}{\updefault}{\color[rgb]{0,0,0}$0010$}%
}}}}
\put(3196,-781){\makebox(0,0)[lb]{\smash{{\SetFigFont{8}{9.6}{\familydefault}{\mddefault}{\updefault}{\color[rgb]{0,0,0}$0001$}%
}}}}
\put(3196,-1231){\makebox(0,0)[lb]{\smash{{\SetFigFont{8}{9.6}{\familydefault}{\mddefault}{\updefault}{\color[rgb]{0,0,0}$0011$}%
}}}}
\put(4321,-1006){\makebox(0,0)[lb]{\smash{{\SetFigFont{8}{9.6}{\familydefault}{\mddefault}{\updefault}{\color[rgb]{0,0,0}$0011$}%
}}}}
\put(4321,-106){\makebox(0,0)[lb]{\smash{{\SetFigFont{8}{9.6}{\familydefault}{\mddefault}{\updefault}{\color[rgb]{0,0,0}$0010$}%
}}}}
\put(-179,-556){\makebox(0,0)[lb]{\smash{{\SetFigFont{8}{9.6}{\familydefault}{\mddefault}{\updefault}{\color[rgb]{0,0,0}$0000$}%
}}}}
\put(6256,-736){\makebox(0,0)[lb]{\smash{{\SetFigFont{8}{9.6}{\familydefault}{\mddefault}{\updefault}{\color[rgb]{0,0,0}$Z$}%
}}}}
\put(6256,-466){\makebox(0,0)[lb]{\smash{{\SetFigFont{8}{9.6}{\familydefault}{\mddefault}{\updefault}{\color[rgb]{0,0,0}$X$}%
}}}}
\put(6256,-331){\makebox(0,0)[lb]{\smash{{\SetFigFont{8}{9.6}{\familydefault}{\mddefault}{\updefault}{\color[rgb]{0,0,0}$I$}%
}}}}
\put(6256,-601){\makebox(0,0)[lb]{\smash{{\SetFigFont{8}{9.6}{\familydefault}{\mddefault}{\updefault}{\color[rgb]{0,0,0}$Y$}%
}}}}
\end{picture}%